\documentstyle[12pt,epsfig]{article}                                
\begin{document}
\titlepage                  
\vspace{1.0in}
\begin{center}
\begin{bf}

\Huge{ Revisited  Swanson's Hamiltonian }

\end{bf}

\vspace{1.0in}

     \Large{ Biswanath Rath }

Department of Physics, North Orissa University, Takatpur, Baripada -757003,
 Odisha, INDIA (e.mail:biswanathrath10@gmail.com)

\end{center} 
\vspace{0.5in}

We notice Swanson's Hamiltonian $ H=w a^{\dagger}a + \alpha a^{2} + \beta (a^{\dagger})^{2} $ , eventhough non-hermitian in nature can allow  real eigenvalues
only  when it is  transformed to equivalent hermitian operator . Further we find real spectra of this oscillator admit only two possible conditions satisfying 
the condition $w\ge \alpha + \beta$  under the relation:$\Omega^{2} = w^{2}-4\alpha\beta$ 


\vspace{0.2in}

Key words-swanson's oscillator , real spectra ,similarity transformation,perturbation theory .

\vspace{0.1cm}
\begin{bf}
I.Introduction
\end{bf}
\vspace{0.1cm}

Since the development of quantum mechanics, it is commonly known that hermitian operators yield real eigenvalues [1]
\begin{equation} 
H  = H^{\dagger}
\end{equation}
However the concept of real spectra took a new turn ,after  Bender and Boettecher[2] proposed a concept: $\mathcal{PT}$ invariant nature on quantum operator.Here $\mathcal{P}$ - operator stands for parity operator and $\mathcal{T}$ stands for time-reversal operator . An operator reflecting $\mathcal{PT}$-invariant  behaviour are non-hermitian in behaviour i.e
\begin{equation} 
H \neq H^{\dagger}
\end{equation}
Probably a simplest model on non-hermitian operator is Swanson's  Hamiltonian[3]
\begin{equation} 
 H = w (a^{\dagger} a + \frac{1}{2})+ \alpha a^{2} + \beta (a^{\dagger})^{2}
\end{equation}
where the parameters $w;\alpha \not = \beta$ are real . Swanson found that spectra of this oscillator are real if and only if 
\begin{equation} 
\Omega^{2} = w^{2} - 4\beta \alpha \gg 0 
\end{equation}
and
\begin{equation} 
 w\gg \alpha+\beta 
\end{equation}

and the energy spectrum is given by 
\begin{equation} 
E_{n}=(n+\frac{1}{2})\Omega
\end{equation}

Under the condition two possible cases arise :

\vspace{0.1cm}

$\bf{case-I}$

\vspace{0.1cm}
For
\begin{equation} 
 w \gg \alpha + \beta 
\end{equation}
the corresponding Hermitian Hamiltonian is [4]
\begin{equation} 
\it{H}=\frac{1}{2}[(w-\alpha - \beta)p^{2} + \frac{(w^{2}-4\alpha\beta)}{(w-\alpha - \beta)} x^{2}]
\end{equation}
whose eigen-spectrum is
\begin{equation} 
E_{n}=\Omega(n+\frac{1}{2})
\end{equation}

$\bf{case-II}$

\begin{equation} 
 w = \alpha + \beta 
\end{equation}
It is interesting to note that many authors [5-10] have studied this oscillator from different angles and none of them has addressed the case-II.
Hence the aim of this paper is to address this point . However before addressing this point we feel to address this system from perturbation theory [11] to  gather some information regarding the constants $\alpha,\beta$ . After gathering the information we will suitably address this system using similarity transformation [4,12].

\vspace{0.1cm}

\begin{bf}
II.Non-Hermitian Perturbation Theory
\end{bf}

\vspace{0.1cm}

Standard perturbation theory [1,11] can be suitable to address Swanson's Hamiltonian if one  incoporates the non-hermitian property of operators . Let 
\begin{equation} 
H=H_{D} + H_{N}
\end{equation}
where $H_{D}$ is 
\begin{equation} 
H_{D}=w(a^{\dagger}a+\frac{1}{2})
\end{equation}
and
\begin{equation} 
H_{N}=\alpha a^{2} + \beta (a^{\dagger})^{2}
\end{equation}
The corresponding eigenvalue relation of unperturbed Hamiltonian is 
\begin{equation} 
H_{D}|\phi_{n}>=[w(a^{\dagger}a+\frac{1}{2})]|\phi_{n}>=(n+\frac{1}{2})|\phi_{n}>
\end{equation}
The energy eigenvalue relation canbe written as [11]
\begin{equation} 
E_{n}=E_{n}^{(0)} + E_{n}^{(2)} + E_{n}^{(4)}+ .........
\end{equation}
Here odd correction terms will be zero . Explicitly
\begin{equation} 
E_{n}^{(0)} = <n|H_{D}|n>_{w}=(n+\frac{1}{2})
\end{equation}
\begin{equation} 
 E_{n}^{(2)} = \sum_{m}\frac{<n|H_{N}|m><m|H_{N}|n>_{w}} {E_{n}^{(0)}-E_{m}^{(0)}}
\end{equation}
\begin{equation} 
 E_{n}^{(4)}=\sum{m,q,r}\frac{<n|H_{N}|m><m|H_{N}|q><q|H_{N}|r><r|H_{N}|n>_{w}}{(E_{n}^{0}-E_{m}^{(0)})(E_{n}^{(0)}-E_{q}^{(0)})(E_{n}^{(0)}-E_{r}^{(0)})}
\end{equation}

On explicit calculation one will see that series becomes convergent if and only if $\frac{\alpha}{w}$;$\frac{\beta}{w}$ must be less than one . If this condition is violated then the infinite series  diverses.

\begin{bf}
III.Similarity Transformation
\end{bf}

 Similarity transformation[4,12] has a specific role in quantum mechanics while dealing with spectral analysis.Many of the spectral properties which can hardly be 
visualised in direct study are easily visualised using similarity transformation. The importance of similarity transformation lies with the problem under investigation . Interestingly problems involving non-hermitian in behaviour are easily studied .
Mathematically let 
\begin{equation} 
 H |\phi>= E|\phi>
\end{equation}
and
\begin{equation} 
 |\Psi>=S |\phi> \hspace{1.0cm} \Longrightarrow \hspace{1.0cm}  <\Psi|=<\phi|S^{-1}
\end{equation}
\begin{equation} 
  S^{-1}S=SS^{-1}=I
\end{equation}
then
\begin{equation} 
SHS^{-1}=h
\end{equation}
Hence it is obvious that 
\begin{equation} 
<\Psi|h|\Psi>=\in=E
\end{equation}
Here wave function $|\phi>$ corresponds to spectrum of the original  hamiltonian $H$ . In fact  spectrum of Hamiltonian $H$ is very difficult to visualise.

\vspace{0.1cm}
\begin{bf}
II.Swanson's Hamiltonian 
\end{bf}
\vspace{0.1cm}

Consider the Hamiltonian
  
Now consider that[12]
\begin{equation} 
 S_{1}=e^{\frac{(a^{\dagger})^{2}}{2}}
\end{equation}
then it is easy to show that 
\begin{equation} 
 h=S_{1}HS_{1}^{-1}= {(w-2\alpha)}(a^{\dagger}a + \frac{1}{2}) \alpha a^{2} + (\alpha + \beta-w)(a^{\dagger})^{2} 
\end{equation}
If $w=\alpha+\beta$ , then the relation can be written as 
\begin{equation} 
 h= {(\beta-\alpha)}(a^{\dagger}a + \frac{1}{2})+ \alpha a^{2} 
\end{equation}

Here  the Hamiltonian is still non-hermitian and energy spectrum is [13]
\begin{equation} 
 \epsilon_{n}=(\beta-\alpha)(n+\frac{1}{2})
\end{equation}
. Now we use a second transformation as[14]

\begin{equation} 
S_{2}=e^{-\alpha \frac{x^{2}}{2(2\alpha-\beta)}} 
\end{equation}

The resulting Hamiltonian[14]

\begin{equation} 
S_{2}hS_{2}^{-1}=h'=\frac{1}{2}[(\beta-2\alpha)p^{2} + 
\frac{(\beta-\alpha)^{2}}{(\beta-2\alpha)}x^{2}]
\end{equation}
One can see that $h'$ is purely hermitian in nature and has the same  energy spectrum
\begin{equation} 
 \in_{n}=(\beta-\alpha)(n+\frac{1}{2})
\end{equation}

\vspace{0.1cm}
\begin{bf}
V.Conclusion
\end{bf}
\vspace{0.1cm}

 We notice that Swanson's Hamiltonian can only yield real spectra if $w\ge \alpha + \beta$. In other words SW under suitable transformation can be converted to Hermitian operator which can be viewed as Harmonic Oscillator under suitable frequency of oscillation .

\vspace{0.1cm}
\begin{bf}
Acknowledgement:
\end{bf}

\vspace{0.1cm}

Author is grateful to Prof Hugh F.Jones ,Imperial College London for his help.

\end{document}